\documentclass[12pt]{iopart}

\usepackage{graphicx}

\begin{document}

\title[IOP Publishing journals]{An alternative transformation factor in the framework of the relativistic aberration of light}

\author{D Roldán$^{1}$\footnote{diego.roldan@ucuenca.edu.ec}, R Sempertegui$^{1}$\footnote{rodrigo.sempertegui@ucuenca.edu.ec}, F Roldán$^{1}$\footnote{francisco.roldan@ucuenca.edu.ec}\\
\small{$^{1}$University of Cuenca}\\
}

\vspace{10pt}
\begin{indented}
\item[]October 2021
\end{indented}

\begin{abstract}
In the present study, we analyze the principles of special relativity and the phenomenon of the aberration of light in combination, deriving a system of equations that allows establishing the relationship between the angles commonly involved in this phenomenon applied within the Minkowski spacetime. As a consequence, an alternative transformation factor (ATF) is obtained that generates two solutions, one of them with the same values as the Lorentz factor. On the other hand, the ATF does not set the speed of light as a limit to the speed of inertial reference frames.
\end{abstract}
%
\vspace{2pc}
\noindent{\it Keywords}: relativistic aberration, alternative Lorentz transformations, faster than light
%
%
%
%

\section{Introduction}
The aberration of light is a phenomenon in astronomy in which the location of celestial objects does not indicate their real position due to the relative speed of the observer and that of light. James Bradley proposed an explanation for this phenomenon as early as 1727, considering the movement of Earth in relation to the Sun \cite{ref01,ref02}. In 1905, Albert Einstein presented an analysis of the aberration of light from the perspective of the special theory of relativity, in the context of which this phenomenon has been investigated through a series of practical and speculative studies with didactic interest \cite{ref03}, with this study falling into the latter category.

The phenomenon of the aberration of light can be described as the angle difference between a light beam in two different inertial reference frames. Relativistic effects can be included in a study to enrich its analysis \cite{ref04}. Thus, for example, the aberration has been related to relativistic Doppler shifts and relativistic velocity addition \cite{ref05}, light-time correction \cite{ref06}, simultaneity \cite{ref07}, Kerr spacetime \cite{ref08}, light refraction \cite{ref09}, gravitational lensing \cite{ref10}, etc.

We carried out a combined study of the phenomenon of the relativistic aberration of light (RAL) and the principles of special relativity (SR), obtaining transformations comparable to the Lorentz transformations (LT) within the Minkowski spacetime. As a consequence of this combination, we obtain a transformation factor very similar to that of the Lorentz factor (LF); these factors largely coincide in terms of their results (calculations), but the new factor presents some additional significant characteristics. According to Dingel et al. \cite{ref11}, when different relativistic effects are combined “it is well-known that richer and bizarre SR behaviors are produced.” In our case, we will just show some results of theoretical interest.
\subsection{An analysis combining the principles of special relativity and the aberration of light}

\begin{figure}
\begin{center}
\includegraphics[width=0.49\textwidth]{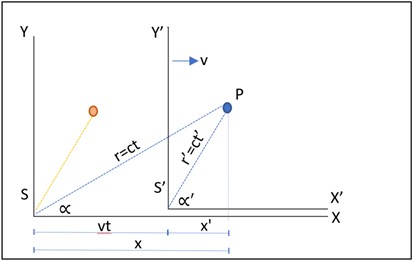}
\end{center}
\caption{Location of a light pulse in relation to two inertial frames.} \label{figure1}
\end{figure}

Figure \ref{figure1} can represent two situations. The first involves (a) a pulse of light that starts from the origin of reference frame S with an angle $\alpha$ and which after a time t is at location P. This same pulse from reference frame S' has an angle $\alpha'$. (b) We can also interpret P as representing an emitter focus that can be perceived from two inertial frames S and S', the first at rest and the second in relative inertial motion. The first scenario is the typical one established to derive the LT, and the second is used to analyze the phenomenon of the RAL. In this paper, we will undertake the exercise of analyzing the two issues together.

Since 1905, many methods of derivation of the LT have been investigated that have enriched our understanding. We undertake a derivation based on the two principles of SR, that is, the principle of relativity and the constancy of the speed of light—or more specifically the “inertial invariance of the speed of propagation of electromagnetic radiation in vacuum” \cite{ref12}—considering the special case described in Figure \ref{figure1}, which implicitly involves the phenomenon of RAL.

\section{Methodology}

It is important to emphasize that our study is limited to the context of the special theory of relativity within the Minkowski spacetime, to all phenomena without significant influence of gravitation. We resort to a thought experiment that is common in studies of SR. We assume two inertial reference frames: S considered at rest and S' with inertial motion and velocity $v$.

We assume that when the origin O' of S' coincides with the origin O of S at the instant $t'=t=0$, a pulse of light is emitted from this origin O with an angle $\alpha$ in the direction of point P. After a time $t>0$, S' will have traveled a distance $vt$ from the perspective of S and $vt'$ from the perspective of S'. This implies that for S', the angle of the trajectory of the light pulse is  $\alpha'$, which due to considerations of the relativistic postulates cannot be derived geometrically directly from Figure \ref{figure1}.

To derive the transformation equations that allow us to convert the measures of S into those of S', the following equations are initially proposed: 


\begin{eqnarray}
x'=\gamma x+\delta t, \label{eq1} \\
t'=\alpha x+\beta t. \label{eq2} 
\end{eqnarray}

The objective is to obtain the values of the four coefficients ($\gamma,\delta,\alpha$, and $\beta$) of equations (\ref{eq1}) and (\ref{eq2}) by identifying special \textit{“known”} cases \footnote{http://web.mst.edu/courses/physics357/Lecture.3.Relativity.Lorentz.Invariance/Lecture3.pdf} concordant with the postulates of special relativity. Each special case is applied to these equations, and subsequently, the values of the coefficients are obtained for general use.

\section{Derivation of relativistic transformations}

\textbf{Special case 1:}

The origin O' from S' will always be at $x'=0$, while for S it will be at $x=vt$.

\begin{eqnarray}
x'=0, \label{eq3} \\
x=vt. \label{eq4} 
\end{eqnarray}

Replacing (\ref{eq3}) and (\ref{eq4}) in (\ref{eq1}):

\begin{eqnarray}
0=\gamma vt+\delta t, \label{eq5} 
\end{eqnarray}

from where
\begin{eqnarray}
\delta =-\gamma v, \label{eq6} 
\end{eqnarray}

Replacing in (\ref{eq1})

\begin{eqnarray}
x'=\gamma x-\gamma vt, \label{eq7} \\
x'=\gamma (x-vt). \label{eq8} 
\end{eqnarray}

By the first postulate, the relationship must be the same (same transformation factor from any of the reference frames). Therefore,
\begin{eqnarray}
x=\gamma(x'+vt' ), \label{eq9} 
\end{eqnarray}

\textbf{Special case 2:}

We can obtain the second special case from the second postulate: $c=c'$.

Generally,

\begin{eqnarray}
r=ct, \label{eq10} \\
r'=ct'. \label{eq11} 
\end{eqnarray}

It should be noted in Figure \ref{figure1} that the path $r>r'$, therefore $ct>ct'$. And since $c$ is a constant, it geometrically suggests that $t>t'$.

Its components on the X-axis are

\begin{eqnarray}
x=cos\alpha\ ct, \label{eq12} \\
x'=cos\alpha'\ ct'. \label{eq13} 
\end{eqnarray}

We isolate from (\ref{eq12}) and (\ref{eq13}) the time variables
\begin{eqnarray}
t=\frac{x}{cos\alpha \ c} \label{eq14} \\
t'=\frac{x'}{cos\alpha' \ c} \label{eq15} 
\end{eqnarray}

We substitute, respectively, in (\ref{eq8}) and (\ref{eq9})
\begin{eqnarray}
x'=\gamma (x-\frac{vx}{cos\alpha\ c})=\gamma x(1 - \frac{v}{cos\alpha \ c}), \label{eq16} \\
x=\gamma (x'+\frac{vx'}{cos\alpha'\ c})=\gamma x'(1+\frac{v}{cos\alpha'\ c}). \label{eq17} 
\end{eqnarray}

We multiply these equations:
\begin{eqnarray}
x'x=\gamma^2 xx' \left(1-\frac{v}{cos\alpha\ c}\right)\left(1+\frac{v}{cos\alpha '\ c}\right) \label{eq18}
\end{eqnarray}

And we obtain
\begin{eqnarray}
\gamma_R=\frac{1}{\sqrt{\left(1-\frac{v}{cos\alpha \ c}\right)\left(+\frac{v}{cos\alpha ' \ c}\right)}} \label{eq19}
\end{eqnarray}

We include the subscript R ($\gamma_R$) to this relativistic transformation factor (\ref{eq19}) derived from aberration concepts to differentiate it from the Lorentz factor (LF) that we refer to with $\gamma_L$. Since this transformation factor includes the aberration angles $\alpha$ and $\alpha'$, we call it the alternative transformation factor (ATF), which differs from the LF traditionally present in the literature. 

Based on the assumptions of this derivation, the ATF can be applied to any aberration angles $\alpha$ and $\alpha'$, whereas LF corresponds only to the special case where $\alpha=\alpha'=0^{\circ}$, which implies that $cos\alpha=cos\alpha'=1$ and thus obtains the LF ($\gamma_L$) usually used in studies on RAL \cite{ref09,ref05}:

\begin{eqnarray}
\gamma_L=\frac{1}{\sqrt{\left(1-\frac{v}{cos0^{\circ}\ c}\right)\left(+\frac{v}{cos0^{\circ}\ c}\right)}}, \label{eq20} \\
\gamma_L=\frac{1}{\sqrt{\left(1-\frac{v^2}{c^2}\right)}}. \label{eq21} 
\end{eqnarray}

As we will see later, the transformation factors (\ref{eq19}) and (\ref{eq21}) generate the same values, so it could be assumed that they are two forms of the same equation, however, equation (\ref{eq19}) allows in some cases values of $v$ greater than those of light, which makes an interesting difference.

\subsection{Relationship between angles of aberration}

A limitation of the ATF (\ref{eq19}) is that we do not yet have a formula for the transformation between the angles $\alpha$ and $\alpha'$. This relationship can be established with the following analysis:

According to Figure \ref{figure1}, in which

\begin{eqnarray}
tan\alpha =\frac{y}{x'} . \label{eq22} 
\end{eqnarray}

since there is no relative motion in the direction of the Y-axis, then
\begin{eqnarray}
y=y'=sin\alpha'\ ct'. \label{eq23} 
\end{eqnarray}

According to (\ref{eq9}),
\begin{eqnarray}
x=\gamma_R (x'+vt'). \label{eq24} 
\end{eqnarray}

Substituting (\ref{eq13}) in (\ref{eq24}),
\begin{eqnarray}
x=\gamma_R (cos\alpha'\ ct'+vt'). \label{eq25} 
\end{eqnarray}

Substituting (\ref{eq23}) and (\ref{eq25}) in (\ref{eq22}),
\begin{eqnarray}
tan\alpha=\frac{sin\alpha'\ ct'}{\gamma_R (cos\alpha'\ ct'+vt') }, \label{eq26} \\ 
tan\alpha=\frac{sin\alpha'}{\gamma_R (cos\alpha' +\frac{vt'}{ct'}) }, \label{eq27} \\ 
tan\alpha=\frac{sin\alpha'}{\gamma_R (cos\alpha' +\frac{v}{c})} . \label{eq28} 
\end{eqnarray}

This equation is remarkably similar to the usual functional form in the relativistic literature, with the difference that the LF is replaced by the new ATF  $\gamma_R$.

If we combine (\ref{eq19}) with (\ref{eq28}), we get
\begin{eqnarray}
\frac{tan\alpha}{\sqrt{1-\frac{v}{cos\alpha\ c}}}=\frac{tan\alpha'}{\sqrt{1+\frac{v}{cos\alpha'\ c}}}. \label{eq29} 
\end{eqnarray}

For comparative purposes, what has been hitherto analyzed is summarized in Table~\ref{math-tab2}.

\begin{table}
\caption{\label{math-tab2}Current system of equations (CSE) and proposed system of equations (PSE) for relativistic aberration.}
\begin{tabular}{ p{4.5cm} p{4.2cm} p{5.8cm} } 
\br
\multicolumn{3}{c}{System of equations:} \\
\hline
&Current (CSE) & Proposed (PSE)\\
\mr
Angle relationships 
& $tan\alpha=\frac{sin\alpha'}{\gamma_L \left(cos\alpha'+\frac{v}{c}\right)}$ 
& $tan\alpha=\frac{sin\alpha'}{\gamma_R \left(cos\alpha'+\frac{v}{c}\right)}$   \\ [17pt]
\hline 
 Transformation factors 
 & $\gamma_L=\frac{1}{\sqrt{\left(1-\frac{v^2}{c^2}\right)}}$ 
 & $\gamma_R=\frac{1}{\sqrt{\left(1-\frac{v}{cos\alpha\ c}\right)\left(+\frac{v}{cos\alpha'\ c}\right)}}$  \\ [15pt]
\br
\end{tabular}
\end{table}

\begin{eqnarray}
\cos\alpha'=\frac{cos\alpha+\frac{v}{c}}{1+\frac{v}{c}cos\alpha}. \label{eq30} 
\end{eqnarray}

It should be noted that in the literature, it is common to find equation (\ref{eq30}) used to establish the relationship between the indicated angles \cite{ref13,ref03}. This equation (\ref{eq30}) provides the same calculations as (\ref{eq21}) and (\ref{eq28}) combined (column CSE Table~\ref{math-tab2}; however, we will use the equations in Table~\ref{math-tab2} because this allows us to analyze the results of the two transformation factors, $\gamma_L$ and $\gamma_R$.

As can be seen, in the CSE the transformation factor corresponds to the LF and does not depend on the angles involved in the RAL, while in the PSE it does. However, as we will discuss below, this does not necessarily lead to different results — at least not always.

Thus, for example, if we consider a velocity $v=0.3c$ and an angle $\alpha'=40^{\circ}$  in S', these data must correspond to an angle $\alpha<\alpha'$ by observation of Figure \ref{figure1}.

According to (\ref{eq21}), 

\begin{equation*}   
\gamma_L=\frac{1}{\sqrt{\left(1-\frac{0.3^2 c^2}{c^2}\right)}}=1.0483
\end{equation*}

According to (\ref{eq28}),
\begin{equation*}
tan\alpha=\frac{sin40^{\circ}}{1.0483(cos40^{\circ} +\frac{0.3c}{c})}=0.57519
\end{equation*}

If we calculate the previous variables using the PSE (equations (\ref{eq19}) and (\ref{eq28})) resorting to numerical analysis, we obtain the same previous values. That is, $\alpha=29.90717^{\circ}$ and $\gamma_R=\gamma_L=1.0483$. 

In general, once the calculations have been made, the two systems of equations provide the same results; this suggests that (\ref{eq21}) can be derived by combining (\ref{eq19}) and (\ref{eq28}); however, the PSE presents some novelties.

\begin{figure}
\begin{center}
\includegraphics[width=0.49\textwidth]{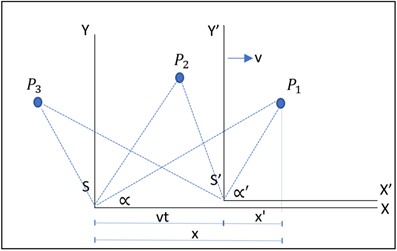}
\end{center}
\caption{Possible trajectories of the light pulse according to the quadrants of the reference frames} \label{figure2}
\end{figure}

The first novelty is that the PSE has two real solutions for the transformation factor $\gamma_R$ in all cases. The first solution always matches the values obtained by the CSE ($\gamma_R=\gamma_L$). The second solution has a behavior that depends on the quadrant in which point P is located, as can be seen in Figure \ref{figure2}. Thus, in the example cited above, in which the velocity $v=0.3c$ and an angle $\alpha'=40^{\circ}$ in S', the first solution was $\alpha=29.90717^{\circ}$ and $\gamma_R=1.0483$, coinciding with the CSE; however, it also corresponds to the values $\alpha=220^{\circ}$ and $\gamma_R=0.718586$. As can be seen, this angle implies that the trajectories are parallel in opposite directions.

In general, when the light pulse is in the first quadrant of S and S' (P1 in Figure \ref{figure2}) or when it is in the second quadrant of S and S' (P3), the second solution corresponds to a \textit{parallel} trajectory in the opposite direction. Speculating on the physical interpretation, these results may suggest that as an effect of relativistic aberration within the Minkowski spacetime, (i.e., even in the absence of gravity), an object in the cosmos can be perceived in two apparent positions at the same time, one of them fully coinciding with the traditional predictions of the SR.

The PSE is indeterminate when the light pulse travels along the Y-axis or the Y'-axis—that is, when the angle $\alpha=90^{\circ}$ or when $\alpha'=90^{\circ}$—since the radicand of the root of the denominator of the factor becomes infinite due to the zero value of the cosine. It should be noted that the above values coincide when $v=-cos\alpha'\ c$ and $v=cos\alpha\ c$, respectively.

On the other hand, when P is in the first quadrant of S and the second of S' (P2), the PSE presents two solutions. As in the previous case, the first also matches the CSE results. Thus, for example, when $v=0.3c$ and $\alpha'=100^{\circ}$, it corresponds to $\alpha=82.3399^{\circ}$ and a transformation factor $\gamma_R=\gamma_L=1.04828$, but it also corresponds to a second solution, $\alpha=80^{\circ}$ and a transformation factor ATF $\gamma_R=1.37432$. In this case, when it comes to P2 the second solution corresponds to supplementary angles ($\alpha+\alpha'=180^{\circ}$). Again, speculatively, these results suggest that as an effect of relativistic aberration a celestial object may have an apparent double location.

\subsection{The time transformation}

For the case of the RAL, we previously obtained the ATF in such a way that we can substitute (\ref{eq19}) in (\ref{eq9}):

\begin{eqnarray}
x=\frac{(x'+vt' )}{\sqrt{(1-\frac{v}{cos\alpha\ c})(1+\frac{v}{cos\alpha'\ c})}}. \label{eq31} 
\end{eqnarray}

Substituting (\ref{eq12}) and (\ref{eq13}) in (\ref{eq31}),
\begin{eqnarray}
cos\alpha\ ct=\frac{cos\alpha'\ ct'+vt'}{\sqrt{(1-\frac{v}{cos\alpha\ c})(1+\frac{v}{cos\alpha'\ c})}}, \label{eq32} \\ 
t=\frac{\frac{cos\alpha'}{cos\alpha}t'+\frac{vt'}{cos\alpha\ c}}
{{\sqrt{(1-\frac{v}{cos\alpha\ c})(1+\frac{v}{cos\alpha'\ c})}}}, \label{eq33} 
\end{eqnarray}

Substituting (\ref{eq15}) in the second term of the numerator, this can also be expressed by 
\begin{eqnarray}
t=\frac{\frac{cos\alpha'}{cos\alpha}t'+\frac{1}{cos\alpha\ cos\alpha'} \frac{vx'}{c^2}}{\sqrt{(1-\frac{v}{cos\alpha\ c})(1+\frac{v}{cos\alpha'\ c})}}. \label{eq34} 
\end{eqnarray}

From the perspective of this derivation equation (\ref{eq34}), like (\ref{eq19}), can be applied to any aberration angles $\alpha$ and $\alpha'$. We can obtain equation (\ref{eq36}), the usual LT of the time, when we consider only the special case where $\alpha=\alpha'=0^{\circ}$, which implies that $cos\alpha=cos\alpha'=1$.
\begin{eqnarray}
t=\frac{(\frac{cos0^{\circ}}{cos0^{\circ}}t'+\frac{1}{cos0^{\circ}cos0^{\circ}}\frac{vx'}{c^2})}{\sqrt{(1-\frac{v}{cos0^{\circ}\ c})(1+\frac{v}{cos0^{\circ}\ c})}}. \label{eq35} \\
t=\frac{(t'+\frac{vx'}{c^2})}{\sqrt{(1-\frac{v^2}{c^2})}}. \label{eq36} 
\end{eqnarray}

\subsection{The case of relativistic aberration of light when $\alpha=180^{\circ}-\alpha'$.}

\begin{figure}
\begin{center}
\includegraphics[width=0.49\textwidth]{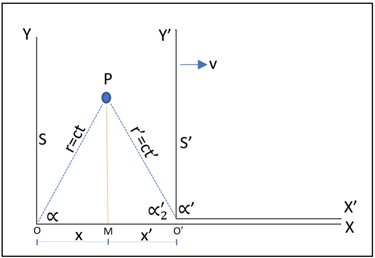}
\end{center}
\caption{Location of a light pulse with trajectories with similar angles ($v=1/2\ cos\alpha\ ct$)} \label{figure3}
\end{figure}

Analyzing Figure \ref{figure3}, it is observed that between P and the origins of S and S' a triangle is formed in which the angles $\alpha$ and $\alpha_2$ depend on the location of P. If P is closer to the Y'-axis, then $\alpha>\alpha_2$; if it is closer to the Y-axis, then $\alpha<\alpha_2$. Therefore, it is feasible to assume that there must exist a location of P where $\alpha=\alpha_2$, which implies that $\alpha'=180^{\circ}-\alpha$. This isosceles triangle actually corresponds to two right triangles (PMO and PMO'), each relative to each frame of reference; however, it allows us to conduct a joint analysis.  

According to the previous assumptions,
\begin{eqnarray}
cos\alpha=-cos\alpha'. \label{eq37} 
\end{eqnarray}

From (\ref{eq33}),
\begin{eqnarray}
t=\frac{t'(\frac{cos\alpha'}{cos\alpha}+\frac{v}{cos\alpha\ c})}
{{\sqrt{(1-\frac{v}{cos\alpha\ c})(1+\frac{v}{cos\alpha'\ c})}}}. \label{eq38} 
\end{eqnarray}

Substituting (\ref{eq37}) in (\ref{eq38}): 
\begin{eqnarray}
t=\frac{t'(-1+\frac{v}{cos\alpha\ c})}{|1-\frac{v}{cos\alpha\ c}|}  . \label{eq39} 
\end{eqnarray}

According to Figure \ref{figure3}, $v>cos\alpha\ c$ (P is between the two Y-axes), so the value of the numerator parentheses of (\ref{eq39}) will be positive. Thus,
\begin{eqnarray}
t=t'. \label{eq40} 
\end{eqnarray}

This result can be explained if we look at Figure \ref{figure3} and remember the usual idea of the light-clock that uses “the distance traveled by a pulse of light and the known speed of light to mark out intervals of time” \cite{ref14}. The common example assumes that $\alpha'=90^{\circ}$, which implies that the greater distance of the path (hypotenuse) of the light pulse corresponds to S. However, any example should be equally valid, and in our case when the two angles are equal the conclusion is that the two paths are also equal ($r=r'$). Therefore, it is intuited that the equality of (\ref{eq40}) is admissible (clock-rate), although only for the case of the aberration of light and when  $\alpha'=180^{\circ}-\alpha$. Obviously, this theoretical result should be supported with empirical evidence. 

\subsection{Faster than the speed of light $v>c$}

An important novelty of the ATF, unlike the LF, is that it allows relative speeds v greater than those of light, although it is not new that in the literature we find some studies that in the context of SR, general relativity, or quantum physics allow us to infer speeds faster than that of light \cite{ref15, ref16}, particularly concerning the case of neutrinos that may be tachyons \cite{ref17, ref18}. 

Thus, for example, according to ATF if the speed of the frame S' is $v=1.2c$ and the angle $\alpha'=100^{\circ}$, an angle $\alpha=80^{\circ}$ corresponds to it—that is, they are supplementary angles—and in this case, $\gamma_R=0.16919$. In this sense, it is possible to speculate a little more about the possibility of hypothetical tachyons \cite{ref19}. It should also be noted that coinciding with Ehrlich, the values of the ATF “are not symmetric about the singularity at $v/c=1$”. These additional characteristics, in agreement with the RAL, determine that the PSE shows a relative anisotropic behavior of the ATF, partially coinciding with the results of Jin and Lazar \cite{ref16}.

\section{Conclusions}

In this paper, we have analyzed, in the context of the Minkowski spacetime, whether the Lorentz factor (LF) used in special relativity should be applied when it comes to the phenomenon of the relativistic aberration of light, concluding that in this case, the alternative transformation factor (ATF) that we derived here in principle provides the same results as the LF; nevertheless, it mathematically allows for second solutions depending on the quadrant in which the emitter focus is located. A particularly important novelty is that the derived ATF allows the reference frames involved to move at speeds greater than that of light.

This study is developed in the context of the relativistic aberration phenomenon, and the system of equations derived includes the angles of the light trajectories as variables. Obviously, it can be assumed that there are other cases of inertial motion that do not involve pulses of light; however, typical relativistic studies cannot dispense with this assumption since it is the only way to mathematize the second postulate of special relativity: the constancy of the speed of light. In this sense, we can ask ourselves if it is feasible for these equations to be extrapolated to a more general context. 

In any case, it must be remembered that a large portion of the empirical evidence that supports the predictions of the LT does not contradict the ATF since the solutions of the former coincide in their mathematical estimates with la latter. However, LF is one of the two possible solutions of the ATF and not just an equivalent mathematical expression. 

Unlike the LF, an important characteristic of the proposed ATF is that it does not prevent the possibility of speeds greater than that of light, offering speculative opportunities for the hypothetical tachyons that are addressed in the literature, as Schwartz [17, 18] does to eventually explain dark energy and dark matter using low energy neutrinos as tachyons. 

Finally, just as aberration distorts the observed position of a celestial body, regardless of the refractive index of a medium [20], the fact that as a consequence of the relativistic aberration the PSE establishes two solutions can be interpreted as the possibility of observing two images of the same celestial body as an effect only of the SR without the influence of gravity of a massive body. These conclusions are limited only to a context of the relativistic aberration under the special relativity approach.
\\
\\


\section*{References}

\begin{thebibliography}{50} 


\bibitem{ref01} T Shivalingaswamy, P. Rashm, \textit{I am the Speed of Light c, You 'see' .....!}, European J of Physics Education \textbf{5} 1, 51-58 (2014).

\bibitem{ref02} J H Sweeney, \textit{Einstein's dreams}, The Review of Metaphysics \textbf{17} 4, 811-834 (2014).

\bibitem{ref03} K Van Acoleyen , J Van Doorsselaere, \textit{Captain Einstein: A VR experience of relativity}, American Journal of Physics, \textbf{88} 10, pp. 801-813, (2020)

\bibitem{ref04} J P Zhu, B Zhang , Y P Yang, \textit{Relativistic Astronomy. II. In-flight Solution of Motion and Test of Special Relativity Light Aberration}, The Astrophysical Journal, \textbf{877} 1, 1-8, (2019). 

\bibitem{ref05} P Saari , I Besieris, \textit{Relativistic aberration and null Doppler shift within the framework of superluminal and subluminal nondiffracting waves}, Journal of Physics Communications, \textbf{4}, 1-10, (2020)

\bibitem{ref06} R R Karimi , D Mortari, \textit{Interplanetary Autonomous Navigation}, Journal of Guidance, Control, and Dynamics, \textbf{38} 6, 1151-56, (2015)


\bibitem{ref07} A Gjurchinovski, \textit{Relativistic aberration of light as a corollary of the relativity of simultaneity}, European Journal of Physics, \textbf{27} 4, 703-708, (2006) 

\bibitem{ref08} H Arakida, \textit{General Relativistic Aberration Equation and Measurable Angle of Light Ray in Kerr Spacetime}, International Journal of Modern Physics D, 2021. 

\bibitem{ref09} A Gjurchinovski, \textit{Aberration of light in a uniformly moving optical medium}, American Journal of Physics, \textbf{72} 7, 934-940, (2004) 

\bibitem{ref10} M Sereno, \textit{Aberration in gravitational lensing}, Physical Review D, \textbf{78} 8, 1-9, (2008) 

\bibitem{ref11} B Dingel, A Buenaventura, A Chua, N Libatique , K Murakawa, \textit{Relativistic aberration of light mimicked by microring resonator based optical All-Pass Filter (APF)}, Optik, \textbf{183} 82-91, (2019) 

\bibitem{ref12} W N Mathews, \textit{Seven formulations of the kinematics of special relativity}, American Journal of Physics, \textbf{88} 269-278, (2020)

\bibitem{ref13} A Jarabo, B Masia, A Velten, C Barsi, R Raskar , D Gutierrez, \textit{Relativistic Effects for Time-Resolved Light Transport}, Computer Graphics Forum, \textbf{34} 8, 1-12, (2015) 

\bibitem{ref14} W B Stannard, \textit{A new model of special relativity and the relationship between the time warps of general and special relativity}, Physics Education, \textbf{53}, 1-9, (2018) 

\bibitem{ref15} J M Hill , B J Cox, \textit{Einstein's special relativity beyond the speed of light}, Proceedings of the Royal Society A, \textbf{468}, 4174–4192, (2012) 

\bibitem{ref16} C Jin, M Lazar, \textit{A note on Lorentz-like transformations and superluminal motion}, Journal of Applied Mathematics and Mechanics, \textbf{95} 7, 690-694, (2015) 

\bibitem{ref17} C Schwartz, \textit{Tachyon dynamics for neutrinos?}, International Journal of Modern Physics A, \textbf{33} 10, 1-23, (2018) 

\bibitem{ref18} C Schwartz, \textit{An approach for modeling tachyons with gravitation}, International Journal of Modern Physics A, \textbf{34} 19, 1-18, (2019) 

\bibitem{ref19} R Ehrlich, \textit{Faster-than-light speeds, tachyons, and the possibility of tachyonic neutrinos}, American Journal of Physics, \textbf{71}, 1109-1114, (2003) 

\bibitem{ref20} N. C. Schmidt, M. Kahms, J. Hüve and J. Klingauf, \textit{FIntrinsic refractive index matched 3D dSTORM with two objectives: Comparison of detection techniques}, Scientific Reports (Nature Publisher Group), 1-12, (2018)  


\end{thebibliography}
\end{document}